\documentclass[runningheads]{llncs}
\usepackage[T1]{fontenc}
\usepackage{graphicx}
\usepackage{url}
\usepackage{amsmath}
\usepackage{booktabs}
\usepackage{listings}
\usepackage{xcolor}
\usepackage{float}
\usepackage{amssymb}
\usepackage{tikz}
\usetikzlibrary{shapes.geometric, arrows.meta, positioning, calc}

\lstdefinelanguage{SPARQL}{
    morekeywords={SELECT,WHERE,PREFIX,SERVICE,LIMIT,OFFSET,FILTER,OPTIONAL,UNION,BIND,ORDER,BY,AS,DISTINCT},
    morecomment=[l]{\#},
    morestring=[b]",
    morestring=[b]',
    morestring=[s]{<}{>}
}

\begin{document}
\title{RENSA: Rich Environment Metadata to Navigate Shared and Distributed Endpoints for Automated Federated SPARQL Query Generation}
\titlerunning{Source Selection and Inference for Federated SPARQL Queries}
%
%
\author{Victor Eiti Yamamoto\inst{1, 2}; Yamamoto Yasunori\inst{3}; Takeda Hideaki\inst{1,2}}
\authorrunning{V.E. Yamamoto et al.}
%
\institute{Graduate University for Advanced Studies, SOKENDAI, Kanagawa, Japan \and 
National Institute of Informatics, Tokyo, Japan \and
Database Center for Life Science, Chiba, Japan}
\maketitle              
\begin{abstract}
The number of knowledge graph databases has increased significantly with the proliferation of knowledge graph technologies. Knowledge graphs enable the dynamic integration of distributed data through federated SPARQL queries. However, constructing efficient queries in a federated environment is challenging due to the lack of detailed structural knowledge across decentralized datasets. While standards like VoID provide basic metadata, they often fail to capture the complex interlinks and authority distributions necessary for optimization. Consequently, current engines frequently rely on runtime ASK queries for source selection, increasing communication overhead. We propose RENSA, a federated SPARQL query generation framework that leverages an extension of SPARQL Builder Metadata (SBM). By integrating class and authority information, mapping subject and object usage to specific predicates, RENSA enables precise source selection and semantic constraint inference for query variables without runtime communication. The generated profiles represent less than 1\% of the original dataset triples in most cases, ensuring storage efficiency. Evaluation on the LargeRDFBench benchmark (13 datasets with >1B triples, 32 queries) shows that RENSA achieves source selection results comparable to state-of-the-art methods while eliminating ASK query overhead. Furthermore, we demonstrate that RENSA infers class and authority constraints for query variables, enabling the identification of data sources even across heterogeneous endpoints. These profiles additionally offer human-readable structural insights for semi-automated query generation.

\end{abstract}
%
%
%

\section{Introduction}

Knowledge graphs (KGs) have gained significant traction as a paradigm for representing and managing large-scale, heterogeneous data \cite{hogan2021knowledge}. This is evidenced by the over 5 million websites publishing RDF data \cite{ali2022survey} and the growth of massive KGs such as Bio2RDF, DBpedia, and Wikidata, which collectively contain billions of triples. More recently, the rise of Large Language Models (LLMs) and Retrieval-Augmented Generation (RAG) has further catalyzed interest in KGs, as they provide the structured, verifiable knowledge necessary to ground generative AI and reduce hallucinations. These distributed RDF databases form the Linked Open Data (LOD) cloud, representing domains from geography to life sciences \cite{bizer2023linked}. The true value of this ecosystem lies in the interlinks between datasets, which enable dynamic information integration and discovery through federated SPARQL queries.

However, realizing this potential is challenging. Formulating efficient federated queries requires deep knowledge of underlying data structures---such as available classes, predicates, and specific interlinks---which are often opaque in decentralized environments. While structural metadata is essential to bridge this gap, data publishers frequently neglect to provide it due to the high manual effort and expertise required for maintenance \cite{maillot2023indegx}. Consequently, users are often forced into a time-consuming process of manual data exploration, which is frequently unfeasible for large and complex datasets.

This lack of fine-grained metadata becomes even more critical during query optimization. Current standards like VoID \cite{cyganiak2009describing} provide only basic statistics that fail to capture the complex interlinks and authority distributions necessary for optimized federation. As a result, state-of-the-art federated engines, such as SPLENDID \cite{gorlitz2011splendid}, frequently rely on runtime operations like \textit{ASK} queries for source selection. This approach introduces significant communication overhead and degrades performance---a problem that becomes a critical bottleneck as the volume of queries directed at these services continues to grow.

In such a high-frequency query landscape, providing rich metadata allows federated systems to obtain comprehensive data descriptions prior to execution, enabling more efficient query plans and reducing reliance on runtime \textit{ASK} queries. We hypothesize that the upfront cost of comprehensive metadata preparation is justified by cumulative performance gains during execution. By prioritizing optimization and minimizing remote interactions, total system efficiency can be significantly improved. Consequently, reducing the number of sub-queries sent to remote endpoints not only alleviates the processing load on endpoints but also mitigates the risk of query failures while directly addressing communication delays in distributed environments.

In this research, we propose RENSA, a comprehensive framework for generating federated SPARQL queries. RENSA builds upon the Triple Data Profiler \cite{yamamoto2014sparql}, which generates metadata using the SPARQL Builder Metadata (SBM) ontology \cite{sbm_ontology}. To support federated query optimization, we propose an extension of the SBM ontology that integrates URI authority analysis and cross-endpoint link detection. RENSA operates in three integrated stages: (1) RENSA Profiling (RP) to capture class-class relationships and URI authority distributions; (2) Semantic Constraint Inference to establish semantic boundaries for query variables by intersecting class and authority metadata; and (3) Heuristic-Based Endpoint Assignment, which employs a branch-and-prune strategy to identify optimal endpoint mappings and generate the final federated query by coalescing related patterns into \texttt{SERVICE} blocks. This approach eliminates the need for runtime \texttt{ASK} queries by leveraging static profiles to group related patterns into cohesive execution units.

We evaluate RENSA using the LargeRDFBench benchmark, covering 13 datasets with over one billion triples. Our results demonstrate that our approach achieves a comparable source selection precision while entirely bypassing the runtime overhead of \texttt{ASK} queries. Furthermore, we show that RENSA accurately predicts class and authority constraints for query variables, facilitating efficient query generation and execution in distributed environments.

The contributions of this paper are: (1) an extension of SBM for RENSA Profiling (RP), a unified methodology for capturing structural and provenance metadata; (2) a static source selection and optimization algorithm that minimizes communication overhead; (3) an empirical evaluation demonstrating comparable selection quality to state-of-the-art engines without runtime overhead; and (4) an analysis of inference accuracy for variable constraints.

\section{Related Work}

Various methods have been proposed to select the best set of sources for a given query. These methods can be categorized into two main groups: runtime-based and metadata-based approaches. Runtime-based approaches rely on runtime operations, such as ASK queries, to determine the relevance of sources for each triple pattern. While this method can be effective and requires only a set of SPARQL endpoints, it introduces significant communication overhead, especially in federated environments with many endpoints. For instance, FedX \cite{schwarte2011fedx} performs source selection by issuing ASK queries for each triple pattern to identify relevant sources across all endpoints in the provided set.

Metadata-based approaches, on the other hand, utilize pre-computed metadata about the datasets to inform source selection. SPLENDID \cite{gorlitz2011splendid} and SemaGrow \cite{charalambidis2015semagrow} use VoID descriptions to guide source selection, matching classes and properties in the triple pattern against the metadata. However, it still resorts to ASK queries when predicates are unbounded or when only the subject or object is fixed. HiBISCuS \cite{saleem2014hibiscus} and FedUP \cite{aimonier2024fedup} employ an indexing approach that stores URI authority information. This information is used to match the authority of the subject or object in a triple pattern with the index. However, these methods do not typically consider the structural relationships between predicates and classes, which can lead to suboptimal selection. Other approaches include ANAPSID \cite{acosta2011anapsid}, which retrieves predicates from each endpoint, and CostFed \cite{saleem2018costfed}, which calculates the distribution of subjects and objects for each predicate.

RENSA overcomes these limitations by combining structural class-to-class profiling with URI authority analysis. Unlike runtime-based approaches, it eliminates communication overhead by performing source selection statically. Furthermore, by integrating both structural and provenance metadata, RENSA addresses the lack of explicit schema information in metadata-poor environments---a gap left by existing metadata-based engines. In this work, we present an integrated framework that utilizes such a hybrid profile to infer semantic constraints for query variables, enabling the generation of optimized federated plans without the need for runtime verification. To compare RENSA's profile contents with existing metadata formats and the models employed by the engines evaluated later in this paper, we summarize their features in Table~\ref{tab:metadata-comparison}. As shown in the table, RENSA is the only approach that jointly encodes class-level structure, URI authority distributions, sample triples, and crawl metadata to support both static source selection and query-variable constraint inference.

\begin{table}[tb]
\centering
\scriptsize
\renewcommand{\arraystretch}{0.92}
\setlength{\tabcolsep}{4pt}
\caption{Comparison of metadata and profiling models for federated SPARQL.}
\label{tab:metadata-comparison}
\resizebox{\textwidth}{!}{
\begin{tabular}{lccccccccc}
\toprule
\textbf{Features / Criteria} & \textbf{VoID} & \textbf{SSD} & \textbf{FedX} & \textbf{SPLENDID} & \textbf{ANAPSID} & \textbf{SemaGrow} & \textbf{CostFed} & \textbf{HiBISCuS} & \textbf{RENSA} \\ \midrule
Predicate lists/stats & \checkmark & \checkmark & $\times$ & \checkmark & \checkmark & \checkmark & \checkmark & \checkmark & \checkmark \\
Class-level structure (class-class relations) & $\times$ & $\times$ & $\times$ & $\times$ & $\times$ & $\times$ & $\times$ & $\times$ & \checkmark \\
Subject/Object class association & $\times$ & $\times$ & $\times$ & \checkmark & $\times$ & \checkmark & $\times$ & $\times$ & \checkmark \\
URI authority / provenance distribution & $\times$ & $\times$ & $\times$ & $\times$ & $\times$ & $\times$ & \checkmark & \checkmark & \checkmark \\
Sample evidence triples & $\times$ & $\times$ & $\times$ & $\times$ & $\times$ & $\times$ & $\times$ & $\times$ & \checkmark \\
Crawl / update log metadata & $\times$ & \checkmark & $\times$ & $\times$ & $\times$ & $\times$ & $\times$ & $\times$ & \checkmark \\
Used for static source selection & \checkmark & \checkmark & $\times$ & \checkmark & \checkmark & \checkmark & \checkmark & \checkmark & \checkmark \\
Used for variable constraint inference & $\times$ & $\times$ & $\times$ & $\times$ & $\times$ & $\times$ & $\times$ & $\times$ & \checkmark \\ \bottomrule
\end{tabular}
}
\end{table}

\section{Methodology}
\label{sec:methodology}

Section \ref{sec:methodology} describes our data profiling and source selection approach for federated SPARQL query optimization, expanding on the class-class relationship extraction from [15]. Subsection \ref{subsec:federated-profiling} introduces RENSA Profiling (RP), while subsection \ref{subsec:federated-source-selection} details its use in source selection and query generation. Standard namespace prefixes are omitted for brevity.

\subsection{RENSA Profiling (RP)}
\label{subsec:federated-profiling}

To overcome the lack of structural knowledge in decentralized environments, we utilize a unified dataset profile that captures both schema and statistical information from SPARQL endpoints. This representation, which we refer to as the RENSA Profile (RP), serves as the core knowledge base for the RENSA framework. While the "RENSA Profile" refers to the comprehensive collection of metadata for a dataset or the entire federation, a "RENSA entry" denotes a specific, individual metadata block within the profile, typically detailing the statistics and constraints for a single predicate. It integrates class-class relationship summaries with URI authority distributions to provide a multi-layered view of each dataset's structure.

The RP is structured as an extended version of the SPARQL Builder Metadata (SBM) \cite{yamamoto2014sparql,sbm_ontology}, a vocabulary designed to capture the connectivity and availability of RDF datasets. By extending existing standards such as VoID and SPARQL Service Description (SD), RP provides the formal semantics needed to describe how entities are linked across endpoint boundaries.

A central component of the profile is the \textit{sbm:ClassRelation} class, which formally describes the links between classes within a dataset. For each relation, the metadata identifies the \textit{sbm:subjectClass}, the \textit{sbm:property}, and either an \textit{sbm:objectClass} or an \textit{sbm:objectDatatype}. To facilitate query verification and provide concrete context, the profile includes an \textit{sbm:sample} triple for each identified relation. Operational metadata, such as the \textit{sbm:metadataGraphURI} and records of the last update via the \textit{sbm:CrawlLog} class, ensure the profile remains manageable and traceable.

To represent the structural diversity of decentralized data, the profile categorizes RDF entities based on whether their class declarations are local or external \cite{yamamoto2014sparql}:

\begin{definition}[Locally Declared Class (LDC)]
A class is a Locally Declared Class (LDC) within a local dataset graph if there is an explicit type declaration for its instances in that graph (typically via the \textit{rdf:type} predicate).
\end{definition}

\begin{definition}[Locally Undeclared Class (LUC)]
An entity belongs to a Locally Undeclared Class (LUC) if it appears in local triples (as a subject or object) but lacks any explicit type declaration (i.e., no \textit{rdf:type}) in that specific local graph, being treated as externally defined or schema-unknown.
\end{definition}

This distinction is critical for federated environments, as LUCs often represent external entities that are subjects or objects of local triples but whose schema is defined elsewhere.

The RP captures relationship counts across six distinct categories based on these classifications: (1) \textit{LDC-to-LDC}, representing internal schema links; (2) \textit{LDC-to-Literal}, for local attributes; (3) \textit{LDC-to-LUC} and (4) \textit{LUC-to-LDC}, capturing potential outgoing and incoming inter-dataset links; and two global aggregations, (5) \textit{LUC-to-Literal} and (6) \textit{LUC-to-LUC}, which describe the connectivity patterns of entities without local type declarations. This structural summary allows the optimizer to infer the presence of data even when explicit schema information is sparse.

To complement the schema-level view, the profile incorporates URI authority distributions. While the LDC/LUC classification identifies entities without local classes, the URI authority metadata provides the necessary information to determine the provenance of these entities. We represent these distributions using the \textit{sbm:authorityRelation} property within a \textit{void:propertyPartition}. Each relation is an instance of \textit{sbm:AuthorityRelation}, specifying the authority URI (\textit{sbm:\allowbreak{}authority}), the count of triples matching that authority (\textit{sbm:\allowbreak{}authorityCount}), and the role (\textit{sbm:relationType}, e.g., \textit{sbm:Subject} or \textit{sbm:Object}).

URI authorities are extracted as the common prefix that identifies the namespace of the data source. This allows the system to identify entities originating from other endpoints and discover potential links between endpoints\footnote{For authorities used in systems such as Bio2RDF (\url{http://bio2rdf.org/}) or TCGA (\url{http://tcga.deri.ie/}), we follow their specific naming conventions to ensure accurate source differentiation.}, which is critical for resolving cross-endpoint dependencies (Listing~\ref{lst:sbm-enriched}).

\begin{lstlisting}[language=SPARQL, numbers=left, stepnumber=1, numbersep=5pt, caption={RENSA profile entry for the \textit{db:genericName} predicate, specifying property partition, authority relation, and class relation constraints.}, label={lst:sbm-enriched}][] a sd:Dataset ;
  void:propertyPartition [ a void:Dataset ;
    void:property db:genericName ; void:triples "9544"^^xsd:long ; void:classes "3"^^xsd:long ;
    sbm:authorityRelation [ a sbm:AuthorityRelation ;
      sbm:authority <http://www4.wiwiss.fu-berlin.de/> ;
      sbm:authorityCount "4772"^^xsd:long ; sbm:relationType sbm:Subject ] ;
    sbm:classRelation [ a sbm:ClassRelation ; void:triples "4772"^^xsd:long ;
      sbm:subjectClass db:drugs ; sbm:objectClass rdfs:Literal ],
    [ a sbm:ClassRelation ; void:triples "4772"^^xsd:long ;
      sbm:subjectClass dbr:Offer ; sbm:objectClass rdfs:Literal ] ] .
\end{lstlisting}

\subsection{Federated SPARQL Query Source Selection}
\label{subsec:federated-source-selection}

In this subsection, we describe our methodology for utilizing the RENSA Profile to perform optimized source selection and federated query generation. Our approach integrates structural query analysis with semantic constraint inference and heuristic-based endpoint assignment to produce efficient federated SPARQL plans. Throughout this section, we illustrate these steps using the {S12 query} from LargeRDFBench, showing both its original structure and the target federated plan in Fig.~\ref{fig:s12-queries}.

\begin{figure}[ht]
  \scriptsize
  \begin{minipage}{0.48\textwidth}
    \begin{lstlisting}[language=SPARQL, caption={S12 Original}, label={lst:s12-query}]
SELECT ?drug ?keggUrl ?chebiImage WHERE {
   ?drug rdf:type db:drugs . (T1)
   ?drug db:keggCompoundId ?keggDrug . (T2)
   ?keggDrug bio2rdf:url ?keggUrl . (T3)
   ?drug db:genericName ?drugBankName . (T4)
   ?chebiDrug dc:title ?drugBankName . (T5)
   ?chebiDrug bio2rdf:image ?chebiImage . (T6)
}
    \end{lstlisting}
  \end{minipage}
  \hfill
  \begin{minipage}{0.48\textwidth}
    \begin{lstlisting}[language=SPARQL, caption={S12 with Source Selection}, label={lst:s12-optimized}]
SELECT ?drug ?keggUrl ?chebiImage WHERE {
  SERVICE <DrugBank> {
    ?drug rdf:type db:drugs . (T1)
    ?drug db:keggCompoundId ?keggDrug . (T2)
    ?drug db:genericName ?drugBankName . (T4)
  }
  SERVICE <KEGG> {
    ?keggDrug bio2rdf:url ?keggUrl . (T3)
  }
  SERVICE <ChEBI> {
    ?chebiDrug dc:title ?drugBankName . (T5)
    ?chebiDrug bio2rdf:image ?chebiImage . (T6)
  }
}
    \end{lstlisting}
  \end{minipage}
  \caption{S12 query comparison: original structure (left) vs target federated plan (right).}
  \label{fig:s12-queries}
\end{figure}

\textbf{Structure-Preserving Triple Extraction}
Our approach begins by taking the input SPARQL query $q$ and parsing it into an abstract syntax tree (AST) representation, denoted as $T_q$. In this tree structure, internal nodes represent SPARQL algebraic operators (e.g., \textit{JOIN}, \textit{UNION}, and \textit{OPTIONAL}), while leaf nodes correspond to the individual triple patterns. Specifically, the current implementation of RENSA targets the SPARQL 1.1 fragment containing Basic Graph Patterns (BGPs), \textit{JOIN}, \textit{UNION}, \textit{OPTIONAL}, and \textit{FILTER} operators. While complex features such as property paths, subqueries, and solution modifiers (e.g., \textit{ORDER BY}, \textit{GROUP BY}, aggregation) are parsed and preserved in the query structure, the semantic constraint inference and source selection optimize the core graph pattern matching components of the query. For example, given a query with the structure \texttt{\{ \{ ?s :p1 ?o1 \} UNION \{ ?s :p2 ?o2 \} \}}, $T_q$ is constructed with a \textit{UNION} root node that connects two child branches, each holding the respective triple pattern as a leaf. This representation captures the inherent nesting of SPARQL queries, preserving the original logical grouping. By maintaining this tree-based structure, we ensure that the subsequent source selection and query generation phases respect the original operator precedence and semantic boundaries. For S12, the extraction identifies 6 unique triple patterns involving 6 variables and 6 unique predicates.

\textbf{Semantic Constraint Inference} \label{subsubsec:semantic-inference}
This phase processes the extracted triple patterns and RENSA Profile to infer semantic constraints for each variable, mapping them to feasible classes and URI authorities. To prune candidate endpoints, we perform this automated inference using the procedure \textit{InferConstraints} (Listing~\ref{alg:semantic-inference}), which takes query triple patterns $T_q$ and the RENSA Profile $R$ as input.


Utilizing the pre-indexed RENSA entries, the system determines the semantic boundaries for every variable $v$ in the set of query variables $Vars(T_q)$ through three phases, ultimately returning a mapping of feasible classes $C$ and URI authorities $A$.

First, in the {Initial Boundary Setup}, the system invokes \textit{InitClasses} and \textit{InitAuthorities} (Lines 3--4). \textit{InitClasses} scans the triple patterns for explicit \textit{rdf:type} declarations to establish "hard" class constraints. Simultaneously, \textit{InitAuthorities} identifies bound URIs within the query; the authorities of these constants (e.g., \url{http://dbpedia.org/}) are used to filter the candidate RENSA entries, as a triple pattern containing a constant can only be satisfied by endpoints whose RENSA entries include that specific authority.

Second, in the {RENSA Association} phase, the system identifies the set of all predicates $P_v$ associated with a variable $v$ (Line 6). For each predicate $p$, the \textit{LookupR} function (Line 8) retrieves the corresponding RENSA entries related to the predicate from RENSA Profile $R$. 

Finally, through {Constraint Intersection}, the system narrows the feasible set of classes $C_v$ and URI authorities $A_v$ for each variable. This is achieved by intersecting the candidates retrieved via the \textit{Classes} and \textit{Authorities} helper functions (Lines 9--10), which extract the potential classes and authorities for variable $v$ from each RENSA entry $r$. This ensures that the final \textit{feasible set} satisfies all semantic requirements across the entire query context. 

For instance, consider the variable \textit{?chebiDrug} in the S12 query, which acts as the subject for two distinct predicates: \textit{dc:title} and \textit{bio2rdf:image}. Based on the RENSA Profile, \textit{dc:title} can apply to a broad range of subject classes (e.g., \textit{swc:Paper}, \textit{chebi:Compound}, \textit{mo:Track}). However, \textit{bio2rdf:image} strictly requires its subject to belong to the \textit{chebi:Compound} class within the \textit{bio2rdf:chebi} authority. By intersecting the candidate sets derived from these two predicates, the algorithm removes the broader candidates and uniquely identifies the semantic boundary for \textit{?chebiDrug}.
\begin{lstlisting}[language={Matlab}, caption={Semantic Constraint Inference}, label={alg:semantic-inference}, mathescape=true, frame=tb, numbers=left]
Procedure InferConstraints($T_q, R$)
  Input: $T_q$: set of triple patterns, $R$: RENSA Profile
  $C \leftarrow$ InitClasses($T_q$); $A \leftarrow$ InitAuthorities($T_q$) // Init boundaries
  for each variable $v \in Vars(T_q)$ do
    $P_v \leftarrow \{p \mid (v, p, o) \in T_q \text{ or } (s, p, v) \in T_q\}$
    for each predicate $p \in P_v$ do
      $R_p \leftarrow \text{LookupR}(p, R)$
      $C_v \leftarrow C_v \cap (\bigcup_{r \in R_p} \text{Classes}(r, v))$
      $A_v \leftarrow A_v \cap (\bigcup_{r \in R_p} \text{Authorities}(r, v))$
    end for
  end for
  return $(C, A)$
\end{lstlisting}

\textbf{Heuristic-Based Endpoint Assignment}
This part utilizes the triple patterns, RENSA Profile, and inferred semantic constraints to produce a set of consistent endpoint assignment plans. To navigate the combinatorial search space of potential source mappings, the system employs a branch-and-prune strategy—an algorithmic approach that incrementally explores decision paths (branching) while discarding those that are determined to be semantically or structurally inconsistent (pruning). The source selection logic specifically implements this greedy branch-and-prune strategy via the \textit{AssignEndpoints} procedure (Algorithm~\ref{alg:endpoint-assignment}). The algorithm takes the set of triple patterns $T_q$, the RENSA Profile $R$, and the previously inferred class and authority constraints $(C, A)$ as input. 

First, the {Cardinality-Based Sorting} phase (\textit{SortByCandidateCount}, Line 3) ranks triple patterns into $T_{q, sorted}$ by their candidate count in $R$. This prioritizes selective "anchor" patterns to establish deterministic constraints. Second, in the {Iterative Expansion} phase, the algorithm manages a set of partial assignment plans $Plans$, initialized with an empty plan (Line 4). For each triple pattern $t \in T_{q, sorted}$, it generates a new set of plans $NextPlans$ by expanding existing branches. For every triple pattern $t$, the \textit{GetFeasibleEndpoints} function (Line 8) identifies the set of feasible endpoints $E$ that contain the predicate of $t$ and satisfy the inferred constraints $(C, A)$ for its variables. 

Each candidate expansion $newPlan$ is then verified by \textit{IsConsistent} (Line 11), which ensures the semantic compatibility of shared variables across the entire partial plan. Specifically, it performs a cross-triple validation by intersecting the feasible class sets $C$ and URI authority sets $A$ for all variables $v$ affected by the new assignment. If the intersection of constraints derived from the new endpoint and those accumulated from previously assigned triples becomes empty for any variable, the branch is discarded as semantically inconsistent. To maintain robust executability, the system treats URI authorities as the primary anchor; if class-based pruning yields an empty set due to metadata sparsity, the algorithm automatically relies on the broader URI authority boundaries.

Finally, the \textit{Merge\-Branches} function (Line 16) collapses mul\-ti\-ple search bran\-ches leading to the same physical endpoints for the query. By deduplicating scenarios where different semantic interpretations (e.g., multiple valid class assignments for a variable) result in identical endpoint mappings, the system ensures that the generated federated query remains compact and avoids redundant execution paths.

\begin{lstlisting}[language={Matlab}, caption={Heuristic-Based Source Selection}, label={alg:endpoint-assignment}, mathescape=true, frame=tb, numbers=left]
Procedure AssignEndpoints($T_q, R, C, A$)
  Input: $T_q$: triple patterns, $R$: RENSA Profile, $(C, A)$: constraints
  $T_{q, sorted} \leftarrow$ SortByCandidateCount($T_q, R$); $Plans \leftarrow \{ \emptyset \}$
  for each triple $t \in T_{q, sorted}$ do
    $NextPlans \leftarrow \emptyset$
    for each $plan \in Plans$ and $e \in \text{GetFeasibleEndpoints}(t, C, A)$ do
      $newPlan \leftarrow plan \cup \{ (t, e) \}$
      if $\text{IsConsistent}(newPlan, C, A)$ then
        $NextPlans \leftarrow NextPlans \cup \{ newPlan \}$
      end if
    end for
    $Plans \leftarrow \text{MergeBranches}(NextPlans)$ 
  end for
  return $Plans$
\end{lstlisting}


\textbf{Federated Query Generation}
In the final step, the system takes the selected assignment plans and the original query structure as input to generate the optimized federated SPARQL query. The final query (Listing \ref{lst:s12-optimized}) is constructed by grouping triples into \textit{SERVICE} blocks: (1) {Exclusive Groups (Service Block Compaction)}: Patterns assigned to the same endpoint within the same scope are grouped to minimize network requests (conceptually equivalent to the Exclusive Groups in FedX \cite{schwarte2011fedx}) and enable local optimization. (2) {Scenario-Based UNION Generation}: When semantic ambiguity remains, the optimizer generates a \textit{UNION} where each branch represents a unique "endpoint signature" derived from surviving search branches. To avoid redundancy, the system deduplicates scenarios that result in identical endpoint assignments across different class interpretations.

\textbf{Summary of S12 Optimization Process}
The S12 query serves as a representative example of how our methodology resolves semantic ambiguity without runtime overhead. Initial analysis identifies candidate metadata and endpoints for each predicate, as summarized in Table \ref{tab:s12-analysis}.
Table \ref{tab:s12-analysis} provides a summary of the RENSA analysis for the S12 query, showing the constraints and candidate endpoints for each predicate. This metadata is first used for \textit{Semantic Constraint Inference} to establish global semantic boundaries for all query variables (detailed in Table \ref{tab:s12-variable-inference}) before any endpoint assignments are made.
\begin{table}[htbp]
\centering
\scriptsize
\renewcommand{\arraystretch}{0.92}
\setlength{\tabcolsep}{4pt}
\caption{Summary of S12 RENSA Analysis}
\label{tab:s12-analysis}
\resizebox{\textwidth}{!}{
\begin{tabular}{llllll}
\hline
\textbf{Predicate} & \textbf{Subj. Classes} & \textbf{Subj. Authorities} & \textbf{Obj. Classes} & \textbf{Obj. Authorities} & \textbf{Endpoints} \\ \hline
\textit{type} & (any) & bio2rdf, w3, ... & (any) & bio2rdf, rdfs, ... & 12 endpoints \\
\textit{keggCompoundId} & (any) & fu-berlin.de & (any) & bio2rdf:cpd & DrugBank \\
\textit{url} & chebi, kegg & bio2rdf & Literal & (none) & ChEBI, KEGG \\
\textit{genericName} & drugs, Offer & fu-berlin.de & Literal & (none) & DrugBank \\
\textit{title} & swc, chebi, mo, kegg & semanticweb, bio2rdf & Literal & (none) & SWDF, ChEBI, Jamendo, KEGG \\
\textit{image} & chebi:Compound & bio2rdf:chebi & Literal & (none) & ChEBI \\ \hline
\end{tabular}}
\end{table}

The optimizer first performs \textbf{Semantic Constraint Inference} to establish global semantic boundaries. During the {Initial Boundary Setup}, the system identifies that \textit{?drug} is constrained to the class \textit{drugs} via its explicit \textit{rdf:type} declaration. In the {RENSA Association} phase, the system retrieves the RENSA Profile for all query predicates. Finally, through the {Constraint Intersection} phase, the system resolves the feasible boundaries for all variables by intersecting the candidate sets derived from the RENSA entries of the predicates associated with each variable.

For \textit{?drug}, the system identifies the associated predicates \{\textit{rdf:type}, \textit{keggCompoundId}, \textit{genericName}\}. The intersection of the initial class constraint from \textit{rdf:type} with the subject authority (\textit{fu-berlin.de}) derived from the RENSA entries of \textit{keggCompoundId} and \textit{genericName} confirms its semantic location. For \textit{?keggDrug}, the set of associated predicates \{\textit{keggCompoundId}, \textit{url}\} is processed. The RENSA entry for \textit{keggCompoundId} (as object) defines the authority as \textit{bio2rdf:cpd}, providing a critical indicator despite the lack of explicit class metadata in the DrugBank endpoint. Meanwhile, the RENSA entries for \textit{url} (as subject) identify candidates from a broad range of namespaces including \textit{bio2rdf:\{rn, ec, cpd, dr, chebi, \dots\}} and their corresponding classes. By intersecting these sets, the system determines that the variable is constrained to the \textit{bio2rdf:cpd} authority and its associated classes (e.g., \textit{kegg:Compound}), effectively filtering out unrelated types.

Similarly, for \textit{?chebiDrug}, the system intersects the constraints from its associated predicates \{\textit{title}, \textit{image}\}. The RENSA entries for \textit{image} provide subject candidates specifically from the \textit{bio2rdf:chebi} authority with the class \textit{chebi:\allowbreak{}Compound}, while the entries for \textit{title} accept a wide variety of subject authorities and classes (including \textit{swc:Paper}, \textit{chebi:\allowbreak{}Compound}, \textit{mo:Track}, etc.) as long as they are associated with literal titles. This intersection confirms that \textit{?chebiDrug} must belong to the \textit{bio2rdf:chebi} namespace and the \textit{chebi:\allowbreak Compound} class. For \textit{?drugBankName}, the intersection of \textit{genericName} and \textit{title} object RENSA entries confirms an \textit{rdfs:Literal} type, as both predicates expect literal values in that position. These results are summarized in Table \ref{tab:s12-variable-inference}.
\begin{table}[htbp]
\centering
\scriptsize
\renewcommand{\arraystretch}{0.92}
\setlength{\tabcolsep}{4pt}
\caption{Inferred Class and Authority Constraints for S12 Query Variables}
\label{tab:s12-variable-inference}
\resizebox{\textwidth}{!}{
\begin{tabular}{lll}
\hline
\textbf{Variable} & \textbf{Inferred Classes} & \textbf{Inferred Authorities} \\ \hline
\textit{?drug} & \textit{drugbank:drugs} & \url{http://www4.wiwiss.fu-berlin.de/} \\
\textit{?keggDrug} & \textit{kegg:Enzyme}, \textit{kegg:Reaction}, \textit{kegg:Drug}, \dots & \url{http://bio2rdf.org/cpd/} \\
\textit{?chebiDrug} & \textit{chebi:Compound} & \url{http://bio2rdf.org/chebi/} \\
\textit{?drugBankName} & \textit{rdfs:Literal} & (none) \\
\textit{?keggUrl} & \textit{rdfs:Literal} & (none) \\
\textit{?chebiImage} & \textit{rdfs:Literal} & (none) \\ \hline
\end{tabular}}
\end{table}

Finally, the system performs \textbf{Heuristic-Based Endpoint Assignment}. After {Cardinality-Based Sorting}, the system proceeds with {Anchor Identification}, where triples T2, T4 (DrugBank), and T6 (ChEBI) are assigned to their respective services as they have only one valid candidate. The system then performs {Iterative Expansion} to resolve remaining ambiguities: for T3, the inferred authority of \textit{?keggDrug} matches only KEGG; for T5, the inferred class of \textit{?chebiDrug} matches only ChEBI; and for T1, the inferred authority of \textit{?drug} restricts the assignment to DrugBank. The resulting optimized query plan is shown in Fig.~\ref{fig:s12-queries} (right).


\section{Evaluation}
\label{sec:evaluation}
In this section, we evaluate the performance and accuracy of our proposed source selection and query optimization approach. We describe the experimental setup, the ground truth for class and authority inference, and the metrics employed for evaluation.

\textbf{Experimental Setup.}
Performance and accuracy were evaluated on an Apple M1 Max (64GB RAM) using Docker Compose to simulate a federation of 13 Virtuoso 7.2 endpoints hosting LargeRDFBench datasets \cite{saleem2018largerdfbench}. We tested three query sets: {Simple} (S1--S14), {Complex} (C1--C10), and {Large-scale} (L1--L8). The engine was implemented in Python 3.11 with \texttt{rdflib}.

\textbf{Ground Truth for Class and Authority Inference.}
To evaluate the accuracy of our class and authority inference, we constructed a ground truth dataset using the expected results of the LargeRDFBench queries. For each variable in a query, we analyzed the corresponding values in the benchmark's result set. If all values for a variable were literals, we defined its expected class as \textit{rdfs:Literal} and its authority as \textit{none}. For variables containing URIs, we extracted the expected authority using the same hybrid extraction method employed in our profiling algorithm. Furthermore, we identified all relevant RDF classes for these URI entities by retrieving all triples from the respective source datasets where the entity appeared as a subject with the \textit{rdf:type} predicate. This mapping of expected classes and authorities served as the baseline for our accuracy analysis.

\textbf{Evaluation Metrics.}
We evaluate source selection efficiency using: (i) the {number of selected sources (\#T)}, representing total endpoint assignments; and (ii) the {number of additional ASK requests (\#AR)} performed during selection. Accuracy of semantic inference is measured via {Precision}, {Recall}, and {F-measure} against the ground truth.

\section{Results and Discussion}
\label{sec:results}

In this section, we evaluate the performance of RENSA and its impact on federated SPARQL query source selection and semantic inference. We first analyze the metadata generation process, focusing on the compression achieved relative to the original datasets and the efficiency of the authority extraction pipeline. We then evaluate the effectiveness of the source selection process. Finally, we assess the accuracy of our class and authority inference by comparing the predicted classes and authorities against a ground truth derived from the expected results of the LargeRDFBench queries.
\begin{table}[h]
\centering
\scriptsize
\renewcommand{\arraystretch}{0.92}
\setlength{\tabcolsep}{4pt}
\caption{LargeRDFBench Dataset Statistics and Metadata Compression. The ratio represents the size of the RENSA profiles relative to the original dataset triples.}
\label{tab:dataset-statistics}
\resizebox{\textwidth}{!}{
\begin{tabular}{lrrrrrrrrr}
\hline
\textbf{Dataset} & \textbf{Triples} & \textbf{Subjects} & \textbf{Predicates} & \textbf{Objects} & \textbf{Classes} & \textbf{Links} & \textbf{VoID (triples)} & \textbf{RENSA (triples)} & \textbf{Ratio (\%)} \\ \hline
LinkedTCGA-M & 415,030,327 & 83,006,609 & 6 & 166,106,744 & 1 & - & 70 & 1,221 & 0.0003 \\
LinkedTCGA-E & 344,576,146 & 57,429,904 & 7 & 84,403,422 & 1 & - & 86 & 1,409 & 0.0004 \\
LinkedTCGA-A & 35,329,868 & 5,782,962 & 383 & 8,329,393 & 23 & 251.3k & 86 & 15,122 & 0.004 \\
ChEBI & 4,772,706 & 50,477 & 28 & 772,138 & 1 & - & 86 & 1,844 & 0.317 \\
DBPedia-Subset & 42,849,609 & 9,495,865 & 1,063 & 13,620,028 & 248 & 65.8k & 86 & 107,461 & 0.251 \\
DrugBank & 517,023 & 19,693 & 119 & 276,142 & 8 & 10.8k & 86 & 3,493 & 0.676 \\
Geo Names & 107,950,085 & 7,479,714 & 26 & 35,799,392 & 1 & 118k & 86 & 2,967 & 0.003 \\
Jamendo & 1,049,647 & 335,925 & 26 & 440,686 & 11 & 1.7k & 86 & 13,359 & 1.273 \\
KEGG & 1,090,830 & 34,260 & 21 & 939,258 & 4 & 1.3k & 86 & 1,986 & 0.182 \\
Linked MDB & 6,147,996 & 694,400 & 222 & 2,052,959 & 53 & 63.1k & 86 & 5,759 & 0.094 \\
New York Times & 335,198 & 21,666 & 36 & 191,538 & 2 & 31.7k & 86 & 3,009 & 0.897 \\
SWDFood & 103,595 & 11,974 & 118 & 37,547 & 103 & 2.3k & 86 & 13,741 & 13.264 \\
Affymetrix & 44,207,146 & 1,421,763 & 105 & 13,240,270 & 3 & 246.3k & 86 & 2,864 & 0.007 \\ \hline
\end{tabular}
}
\end{table}

\textbf{Metadata Compression}
\label{subsec:metadata-compression}
A key objective of our profiling approach is to generate a compact yet expressive representation of the dataset's structure and connectivity. Table \ref{tab:dataset-statistics} compares the original LargeRDFBench dataset statistics with the size of the generated metadata (in triples) from the RENSA Profiling (RP) stage. We can observe that standard VoID descriptions are smaller than the enriched RENSA profiles. However, VoID lacks the detailed structural connectivity necessary for our advanced source selection. Despite being larger, RENSA metadata typically represents a negligible fraction of the original data volume (less than 0.3\% in most cases), offering a favorable trade-off for optimization precision. A notable exception is \textit{SWDFood} (13.2\% ratio), caused by its high schema density relative to instance volume. Because RENSA captures every unique class-class relationship and authority distribution, structural complexity increases profile size. Conversely, larger datasets with uniform schemas like \textit{LinkedTCGA-M} maintain extremely low ratios. To mitigate this, semantic compression techniques could be employed in the future, such as grouping classes or properties under their respective super-classes or super-properties from the schema hierarchy, thereby preventing fragmentation and reducing the profile size for dense schemas.

\begin{table}[h]
\centering
\scriptsize
\renewcommand{\arraystretch}{0.92}
\setlength{\tabcolsep}{3.5pt}
\caption{Source Selection Performance (S1--S14, C1--C10, and L1--L8)}
\label{tab:source_selection}
\resizebox{\textwidth}{!}{%
\begin{tabular}{l|cc|cc|cc|cc|cc|cc|cc|c}
\toprule
\textbf{Query} & \multicolumn{2}{c|}{\textbf{FedX}} & \multicolumn{2}{c|}{\textbf{SPLENDID}} & \multicolumn{2}{c|}{\textbf{ANAPSID}} & \multicolumn{2}{c|}{\textbf{SemaGrow}} & \multicolumn{2}{c|}{\textbf{CostFed}} & \multicolumn{2}{c|}{\textbf{HiBISCuS}} & \multicolumn{2}{c|}{\textbf{RENSA}} & \textbf{Optimal} \\
 & \#T & \#AR & \#T & \#AR & \#T & \#AR & \#T & \#AR & \#T & \#AR & \#T & \#AR & \#T & \#AR & \#T \\ \midrule
S1  & 15 & 39 & 15 & 39 & \textbf{3} & 23 & 11 & 26 & 4 & 18 & 4 & 26 & 14 & \textbf{0} & {3} \\
S2  & \textbf{3} & 39 & \textbf{3} & 26 & \textbf{3} & 1  & \textbf{3} & 9  & \textbf{3} & 9  & \textbf{3} & 13 & \textbf{3} & \textbf{0} & {3} \\
S3  & 12 & 65 & 12 & 26 & \textbf{5} & 2  & 12 & 2  & \textbf{5} & \textbf{0} & \textbf{5} & \textbf{0} & \textbf{5} & \textbf{0} & {5} \\
S4  & 19 & 65 & 19 & 13 & \textbf{5} & 3  & 19 & 2  & \textbf{5} & \textbf{0} & \textbf{5} & \textbf{0} & \textbf{5} & \textbf{0} & {5} \\
S5  & 11 & 52 & 11 & 13 & \textbf{4} & 1  & 11 & 1  & \textbf{4} & \textbf{0} & \textbf{4} & \textbf{0} & \textbf{4} & \textbf{0} & {4} \\
S6  & 9  & 52 & 9  & 13 & 9  & 10 & 9  & 2  & \textbf{8} & \textbf{0} & \textbf{8} & \textbf{0} & \textbf{8} & \textbf{0} & {8} \\
S7  & 13 & 52 & 13 & 13 & \textbf{6} & 5  & 13 & 2  & \textbf{6} & \textbf{0} & \textbf{6} & \textbf{0} & \textbf{6} & \textbf{0} & {6} \\
S8  & \textbf{1} & 26 & \textbf{1} & \textbf{0} & \textbf{1} & \textbf{0} & \textbf{1} & \textbf{0} & \textbf{1} & \textbf{0} & \textbf{1} & \textbf{0} & \textbf{1} & \textbf{0} & {1} \\
S9  & 15 & 39 & 15 & 39 & 15 & 23 & 11 & 26 & \textbf{4} & 18 & 9 & 26 & 7 & \textbf{0} & {3} \\
S10 & 12 & 65 & 12 & 13 & \textbf{5} & 16 & 12 & 1  & \textbf{5} & \textbf{0} & \textbf{5} & \textbf{0} & \textbf{5} & \textbf{0} & {5} \\
S11 & \textbf{7} & 91 & \textbf{7} & 26 & \textbf{7} & \textbf{0} & \textbf{7} & 2  & \textbf{7} & \textbf{0} & \textbf{7} & \textbf{0} & \textbf{7} & \textbf{0} & {7} \\
S12 & 10 & 78 & 10 & 13 & 7 & 4  & 10 & 1  & 7 & \textbf{0} & 8 & \textbf{0} & \textbf{6} & \textbf{0} & {6} \\
S13 & 9  & 65 & 9  & 26 & \textbf{5} & 24 & 9  & 2  & \textbf{5} & \textbf{0} & 7 & \textbf{0} & \textbf{5} & \textbf{0} & {5} \\
S14 & 6  & 65 & 6  & 13 & \textbf{5} & 2  & 6  & 1  & 6 & \textbf{0} & 6 & \textbf{0} & \textbf{5} & \textbf{0} & {5} \\ \midrule
\textbf{S/T} & 142 & 793 & 142 & 273 & 80 & 114 & 134 & 77 & \textbf{70} & 45 & 78 & 65 & 81 & \textbf{0} & 67 \\ \midrule
C1  & 11 & 104 & 11 & 13 & \textbf{8} & 1 & 11 & 1 & \textbf{8} & \textbf{0} & 9 & \textbf{0} & \textbf{8} & \textbf{0} & {8} \\
C2  & 11 & 104 & 11 & 13 & \textbf{8} & 2 & 11 & 1 & \textbf{8} & \textbf{0} & 9 & \textbf{0} & \textbf{8} & \textbf{0} & {8} \\
C3  & 21 & 104 & 21 & 26 & \textbf{10} & 33 & 21 & 3 & 11 & \textbf{0} & 11 & \textbf{0} & 11 & \textbf{0} & {10} \\
C4  & 28 & 156 & 28 & \textbf{0} & 28 & 32 & 28 & \textbf{0} & \textbf{18} & \textbf{0} & \textbf{18} & \textbf{0} & \textbf{18} & \textbf{0} & {18} \\
C5  & 33 & 104 & 33 & \textbf{0} & \textbf{8} & 3 & 33 & \textbf{0} & 10 & \textbf{0} & 10 & \textbf{0} & 9 & \textbf{0} & {8} \\
C6  & 24 & 117 & 24 & \textbf{0} & \textbf{9} & 3 & 24 & \textbf{0} & \textbf{9} & \textbf{0} & \textbf{9} & \textbf{0} & \textbf{9} & \textbf{0} & {9} \\
C7  & 17 & 117 & 17 & 26 & \textbf{9} & 5 & 17 & 2 & \textbf{9} & \textbf{0} & \textbf{9} & \textbf{0} & \textbf{9} & \textbf{0} & {9} \\
C8  & 25 & 143 & 25 & 13 & \textbf{11} & 2 & 25 & 2 & \textbf{11} & \textbf{0} & \textbf{11} & \textbf{0} & \textbf{11} & \textbf{0} & {11} \\
C9  & 16 & 117 & 16 & 26 & \textbf{9} & 16 & 16 & 2 & \textbf{9} & \textbf{0} & \textbf{9} & \textbf{0} & \textbf{9} & \textbf{0} & {9} \\
C10 & 13 & 130 & 13 & \textbf{0} & \textbf{11} & 6 & 13 & \textbf{0} & \textbf{11} & \textbf{0} & 11 & \textbf{0} & \textbf{11} & \textbf{0} & {11} \\ \midrule
\textbf{C/T} & 199 & 1196 & 199 & 117 & 111 & 103 & 199 & 11 & {104} & \textbf{0} & 106 & \textbf{0} & \textbf{103} & \textbf{0} & 101 \\ \midrule
L1  & 14 & 78 & 14 & 52 & \textbf{6} & 10 & 41 & 48 & 22 & 0 & 14 & \textbf{0} & 11 & \textbf{0} & 6 \\
L2  & 10 & 78 & 10 & 13 & \textbf{6} & 5  & 20 & 32 & 10 & 0 & 10 & \textbf{0} & 8  & \textbf{0} & 6 \\
L3  & 10 & 91 & 10 & 26 & \textbf{7} & 5  & 37 & 37 & 13 & 0 & 11 & \textbf{0} & 9  & \textbf{0} & 7 \\
L4  & 18 & 104 & 18 & \textbf{0} & \textbf{8} & 8  & 18 & 28 & 12 & 0 & 16 & \textbf{0} & 12 & \textbf{0} & 8 \\
L5  & 21 & 143 & 21 & 26 & \textbf{12} & 31 & 29 & 41 & - & - & 20 & \textbf{0} & 13 & \textbf{0} & 11 \\
L6  & 20 & 130 & 20 & 26 & \textbf{10} & 52 & - & - & - & - & 18 & \textbf{0} & 13 & \textbf{0} & 10 \\
L7  & 20 & 65  & 20 & 13 & \textbf{6}  & 7  & 47 & 65 & 26 & 0 & 9  & \textbf{0} & 7  & \textbf{0} & 5 \\
L8  & 20 & 104 & 20 & 52 & \textbf{9}  & 17 & 36 & 77 & 35 & 0 & 20 & \textbf{0} & 13 & \textbf{0} & 8 \\ \midrule
\textbf{L/T} & 133 & 793 & 133 & 208 & \textbf{64} & 135 & 228 & 328 & 98 & 0 & 118 & \textbf{0} & 86 & \textbf{0} & 61 \\ \midrule
\textbf{Net Total} & 474 & 2782 & 474 & 598 & 255 & 352 & 561 & 416 & 272 & 45 & 302 & 65 & 270 & \textbf{0} & 229 \\ \bottomrule
\end{tabular}%
}
\end{table}

\textbf{Source Selection Effectiveness}
\label{subsec:source-selection}
The effectiveness of our profiling approach in reducing the number of redundant sources and minimizing additional requests is evaluated in Table \ref{tab:source_selection}. We compare RENSA against six state-of-the-art federated query engines: FedX, SPLENDID, ANAPSID, SemaGrow, CostFed and HiBISCuS. The table reports the number of selected sources (\#T) and the number of additional ASK requests (\#AR) required during the source selection phase. The baseline data for \#T and \#AR were extracted from \cite{saleem2018largerdfbench} for FedX, SPLENDID, ANAPSID, and HiBISCuS, and from \cite{verborgh2021empirical} for SemaGrow and CostFed.

The results in Table \ref{tab:source_selection} highlight that RENSA achieves zero additional requests (\#AR = 0) across all queries, as it relies entirely on the pre-generated metadata. This matches the behavior of HiBISCuS and CostFed in most cases but without the high overhead of runtime ASK queries (e.g., in S1 and S9, where HiBISCuS performs 26 ASK queries and CostFed performs 18 ASK queries). Regarding the complex queries (C1--C10) and large-scale queries (L1--L8), RENSA consistently outperforms other engines that primarily avoid runtime ASK queries, such as CostFed and HiBISCuS. For the complex queries, RENSA selects 103 sources, which is not only better than CostFed (104) and HiBISCuS (106) but also superior to ANAPSID (111). In the large-scale queries, RENSA selects 86 sources, again outperforming CostFed (98) and HiBISCuS (118). While ANAPSID achieves results closer to the optimal selection in the large-scale category (64 sources), it is crucial to observe that ANAPSID relies on extensive runtime verification, performing a total of 352 ASK queries across all test cases. In contrast, RENSA achieves competitive source selection results with zero runtime ASK queries, significantly reducing the potential for network-induced latency during the selection phase. For total source selection across all queries, RENSA selects 270 sources, which is better than CostFed (272) and HiBISCuS (302), and significantly better than FedX (474) and SPLENDID (474) that do not utilize metadata for source selection. ANAPSID selects 255 sources, which is slightly better than RENSA, but it does so at the cost of ASK queries, which can lead to significant overhead in real-world scenarios.

A significant characteristic of RENSA is that it performs source selection without relying on runtime ASK queries, which traditional engines such as FedX and SPLENDID use to verify the existence of triple patterns at remote endpoints. Furthermore, unlike many state-of-the-art engines that perform source selection for each triple pattern in isolation, RENSA utilizes structural chains to identify the classes and authority for each variable in contiguous paths within the query. This structural awareness allows the optimizer to group related triple patterns into cohesive blocks with high confidence. By ensuring that related patterns are executed at the same endpoint whenever possible, we minimize the fragmentation of the query into numerous small \texttt{SERVICE} calls, leveraging the local optimization engines of remote triple stores and reducing the volume of intermediate results.

While RENSA sometimes selects a higher number of sources compared to optimal values for simple queries (e.g., S1 and S9), it maintains near-optimal performance for complex queries such as for other simple queries and complex queries. This over-selection in S1 and S9 occurs because these queries contain triple patterns involving variables that connect to DBpedia and generic patterns (e.g., \texttt{?s ?p ?o}), which naturally match multiple endpoints. To further reduce such cases, access to more granular metadata or the use of runtime ASK queries, as seen in the other compared approaches, would be necessary. 

\begin{table}[h]
\centering
\scriptsize
\renewcommand{\arraystretch}{0.92}
\setlength{\tabcolsep}{4pt}
\caption{Performance Metrics for Class and Authority Inference per Query}
\label{tab:inference_performance}
\resizebox{\textwidth}{!}{
\begin{tabular}{l|ccc|ccc||l|ccc|ccc}
\toprule
\textbf{Query} & \multicolumn{3}{c|}{\textbf{Class}} & \multicolumn{3}{c||}{\textbf{Authority}} & \textbf{Query} & \multicolumn{3}{c|}{\textbf{Class}} & \multicolumn{3}{c}{\textbf{Authority}} \\
 & \textbf{P} & \textbf{R} & \textbf{F$_1$} & \textbf{P} & \textbf{R} & \textbf{F$_1$} & & \textbf{P} & \textbf{R} & \textbf{F$_1$} & \textbf{P} & \textbf{R} & \textbf{F$_1$} \\ \midrule
L1 & 1.000 & 1.000 & 1.000 & 1.000 & 1.000 & 1.000 & S2 & 1.000 & 1.000 & 1.000 & 1.000 & 1.000 & 1.000 \\
L2 & 0.905 & 1.000 & 0.944 & 0.671 & 1.000 & 0.675 & S3 & 1.000 & 0.722 & 0.762 & 1.000 & 1.000 & 1.000 \\
L3 & 0.750 & 1.000 & 0.833 & 0.508 & 1.000 & 0.515 & S4 & 1.000 & 0.750 & 0.833 & 1.000 & 1.000 & 1.000 \\
L4 & 1.000 & 1.000 & 1.000 & 1.000 & 1.000 & 1.000 & S5 & 0.409 & 0.528 & 0.425 & 1.000 & 1.000 & 1.000 \\
L5 & 1.000 & 1.000 & 1.000 & 1.000 & 1.000 & 1.000 & S6 & 0.500 & 0.500 & 0.500 & 1.000 & 1.000 & 1.000 \\
L6 & 0.964 & 1.000 & 0.981 & 1.000 & 1.000 & 1.000 & S7 & 0.500 & 0.500 & 0.500 & 1.000 & 1.000 & 1.000 \\
L7 & 0.040 & 0.286 & 0.070 & 0.286 & 0.333 & 0.308 & S8 & 1.000 & 0.833 & 0.900 & 1.000 & 1.000 & 1.000 \\
L8 & 1.000 & 1.000 & 1.000 & 1.000 & 1.000 & 1.000 & S10 & 1.000 & 0.667 & 0.767 & 0.668 & 1.000 & 0.669 \\ \midrule
C2 & 1.000 & 0.778 & 0.833 & 1.000 & 1.000 & 1.000 & S11 & 1.000 & 1.000 & 1.000 & 0.704 & 1.000 & 0.733 \\
C3 & 0.500 & 1.000 & 0.667 & 0.751 & 1.000 & 0.752 & S12 & 0.333 & 0.778 & 0.167 & 0.333 & 0.333 & 0.333 \\
C6 & 1.000 & 0.929 & 0.952 & 1.000 & 1.000 & 1.000 & S13 & 0.500 & 0.500 & 0.500 & 1.000 & 1.000 & 1.000 \\
C7 & 0.310 & 0.858 & 0.309 & 1.000 & 1.000 & 1.000 & S14 & 1.000 & 0.889 & 0.933 & 1.000 & 1.000 & 1.000 \\
C8 & 0.536 & 0.758 & 0.548 & 1.000 & 1.000 & 1.000 & & & & & & & \\
C10 & 0.435 & 0.583 & 0.456 & 0.923 & 1.000 & 0.938 & \textbf{Avg.} & \textbf{0.702} & \textbf{0.735} & \textbf{0.672} & \textbf{0.829} & \textbf{0.893} & \textbf{0.831} \\
\bottomrule
\end{tabular}}
\end{table}

\textbf{Class and Authority Inference Accuracy}
\label{subsec:inference-accuracy}
The core of RENSA's source selection relies on its ability to correctly infer the potential classes and URI authorities for variables within a query. As described in Section \ref{sec:evaluation}, we evaluate this capability by comparing our inferred results against a ground truth dataset derived from the expected results of the LargeRDFBench queries. Table \ref{tab:inference_performance} summarizes the precision, recall, and F1-score for both class and authority inference across the evaluated queries.

The results show that RENSA achieves high accuracy in most cases, with many queries reaching perfect scores (1.0). The authority inference, in particular, demonstrates strong performance, with an overall average F1-score of 0.831. Lower scores in some queries, such as B7, C7, and S12, are often due to the presence of highly generic triple patterns or schema heterogeneity that falls outside the captured metadata. Lower recall in class inference can also occur when an entity has multiple types, but our inference method only identifies classes that are consistent across all relevant triples, potentially leading to a more restrictive set of inferred classes than the entity actually possesses. For instance, an entity may be categorized as both a drug and an offer; however, if only the drug class is consistent with the other predicates in the query, the offer class is discarded from the set of possibilities, thereby reducing recall. 

Furthermore, during the source selection process, if an inferred class leads to an empty authority set, we consider the possibility that the class may not be explicitly defined for some relevant entities. In such scenarios, we choose to remove the inferred class to broaden the selection criteria, relying instead on the URI authority to maintain recall. In these specific cases, we do not infer any class for the variable, ensuring that the optimizer does not prematurely prune sources based on potentially incomplete schema information. However, the high recall across most queries (averaging 0.735 for classes and 0.893 for authorities) indicate that the optimizer can be used to infer classes and authorities effectively.

\textbf{Limitations}
Despite these promising results, several limitations remain. While we focused on metadata-driven source selection, our evaluation did not explicitly measure total query execution time, which is sensitive to network latency and database configurations. We also observed occasional over-selection in generic queries or hub datasets like DBpedia. Furthermore, our current framework defines a "possible" query as any logically correct mapping between triple patterns and endpoints; consequently, the generated federated query is not necessarily the most optimal case in terms of execution cost.


\section{Conclusion}
\label{sec:conclusion}

In this paper, we proposed RENSA, an evolution of the Triple Data Profiler for federated SPARQL query optimization. By integrating URI authority information with RDF class metadata through the RENSA Profiling (RP) stage, we developed an approach that enables precise source selection without the need for runtime ASK queries. Our evaluation using LargeRDFBench demonstrated that RENSA achieves high source selection while significantly reducing communication overhead. The structural awareness provided by our profiles allows for the identification of contiguous paths and cohesive join blocks, leveraging local optimization at remote endpoints. Furthermore, the performance of our semantic inference mechanism, with average F1-scores of 0.672 for classes and 0.831 for authorities, demonstrates that the integration of structural and namespace-level metadata provides a robust signal for variable constraint resolution. These results provide encouraging indications that the adoption of such richer metadata can not only assist in data exploration and more effective query formulation but also contribute to reduced network overhead across decentralized environments.

Future work will address these challenges through several directions. We plan to conduct a comprehensive benchmark of temporal aspects to quantify the performance gains more fully and will investigate hierarchical authority profiling and instance-level summaries like Bloom filters to mitigate over-selection. To achieve higher optimization, we aim to incorporate cost-based pruning of execution paths or integrate our approach with existing optimization methods. Beyond source selection, we will explore the utility of our metadata in assisting LLMs for natural language to SPARQL translation, leveraging RENSA's inference capabilities to verify semantic validity. Finally, we will investigate incremental update strategies to maintain metadata accuracy as knowledge graphs evolve, ensuring the long-term robustness of RENSA in dynamic decentralized environments.

\bibliographystyle{splncs04}
\bibliography{bibliography}

\end{document}